\documentclass[aps,prd,twocolumn,groupedaddress,showpacs,nofootinbib,amssymb]{revtex4-2}
\usepackage[dvips]{graphicx}
\usepackage{amssymb}
\usepackage{amsmath}
\usepackage{graphicx,,color}
\usepackage{amsfonts}
\usepackage{bm}
\usepackage{microtype}
\usepackage{cancel}
\usepackage{comment}

\newcommand\be{\begin{equation}}
\newcommand\ee{\end{equation}}

\allowdisplaybreaks[4]

\begin{document}

\tolerance=5000

\title{Late-time Cosmology of scalar field assisted $f(\mathcal{G})$ gravity}

\author{
S.A.
Venikoudis,\,\thanks{venikoudis@gmail.com}
K.V. Fasoulakos,\,
\thanks{kvfasoulakos@gmail.com}
F.P.
Fronimos\, \thanks{fotisfronimos@gmail.com}}
\affiliation{
 Department of Physics, Aristotle University of
Thessaloniki, Thessaloniki 54124,
Greece\\}

\tolerance=5000

\begin{abstract}
In this work we present the late-time behaviour of the Universe in the context of Einstein-Gauss-Bonnet gravitational theory. The theory involves a scalar field, which represents low-effective quantum corrections, assisted by a function $f(\mathcal{G})$ solely depending from the Gauss-Bonnet topological invariant $\mathcal{G}$. It is considered that the dark energy serves as the impact of all geometric terms, which are included in the gravitational action and the density of dark energy acts as a time dependent cosmological constant evolving with an infinitesimal rate and driving the Universe into an accelerating expansion. We examine two cosmological models of interest. The first involves a canonical scalar field in the presence of a scalar potential while the second, involves a scalar field which belongs to a generalized class of theories $f(\phi,X)$ namely the k-essence scalar field in the absence of scalar potential. As it is proved, the aforementioned models are in consistency with the latest Planck data and in relatively good agreement with the $\Lambda$CDM standard cosmological model. The absence of dark energy oscillations at the early stages of matter dominated era, which appear in alternative scenarios of cosmological dynamics in the context of modified $f(R)$ gravitational theories, indicates an advantage of the theory for the interpretation of late-time phenomenology.
\end{abstract}


\maketitle

\section{Introduction}
Observations on Supernovae type Ia \cite{SupernovaSearchTeam:1998fmf} have shown that
the Universe expands with an accelerating rate. However, the cosmological dynamics during the late-time era of our Universe remains one of the most intriguing mystery for modern theoretical cosmologists until today. The most successful interpretation for the cosmological evolution is given by the 
$\Lambda$CDM model, which is the most well-established theory in consistency with the observations from the data of Cosmic Microwave Background. In the context of the aforementioned model, $\Lambda$ serves as a cosmological constant, which dominates in late-time era compared with the relativistic and non-relativistic matter. However, several problems have yet to be solved with the most important of them, the understanding of the nature of the cosmological constant, which is related with the dark energy. In addition, based on the Einstein-Hilbert gravitational theory the equation of state parameter of $\Lambda$ is considered that remains constant, but there is lack of certainty from observational perspective.

Beyond the Einstein's gravity many modified gravitational theories have been arisen as extensions, in order to provide a solid explanation of the late-time phenomenon, see Ref. \cite{Oikonomou:2020oex,Oikonomou:2020qah,Odintsov:2019evb,Oikonomou:2013rba,Odintsov:2017hbk,Nojiri:2019fft,Sahni:2014ooa,Nojiri:2017ncd,Nojiri:2009kx,Capozziello:2011et,Nojiri:2006ri,Nojiri:2010wj,Olmo:2011uz,Nojiri:2003ft,Nojiri:2007as,Nojiri:2007cq,Cognola:2007zu,Nojiri:2006gh,Appleby:2007vb,Zhong:2018tqn,Li:2007jm,Nojiri:2005jg,Nojiri:2005am,Cognola:2006eg,Elizalde:2010jx,Izumi:2014loa,Oikonomou:2016rrv,Kleidis:2017ftt,Escofet:2015gpa,Makarenko:2017vuk,Makarenko:2016jsy,Navo:2020eqt,Bajardi:2020osh,Capozziello:2019wfi,Benetti:2018zhv,Clifton:2006kc,Barrow:1988xh,Bamba:2009uf,DeLaurentis:2015fea,DeFelice:2010sh,delaCruzDombriz:2011wn,Elizalde:2010ts,Odintsov:2020nwm}. These theories involve higher order gravitational terms which originate from curvature invariants. The main advantage of these theories is that, they can provide unification models between the post quantum primordial era namely, the inflationary era and the late-time era compatible with the latest Planck data and with the $\Lambda$CDM model. On the other hand, $f(R)$ theories have a serious drawback under certain circumstances. As the numerical value of redshift increases, dark energy oscillations seams to appear in several models, for instance power-law models with an exponent $n$ satisfying the condition $0<n<1$, specifically during the last stages of the matter domination \cite{Bamba:2012qi}. These oscillations originate from the existence of higher derivatives of the Hubble rate. If one uses nontrivial models such as singular power-law models $f(R)$ with exponents $n\leq-1$ oscillations seem to disappear, see Ref \cite{Fronimos:2021ejo}.

In this paper it is demonstrated the phenomenology of the late-time Universe of a given scalar-tensor theory in the context of string-inspired gravity, see Ref. \cite{Oikonomou:2021kql,Odintsov:2020sqy,Oikonomou:2020sij,Oikonomou:2020tct,Oikonomou:2020tct,Odintsov:2020ilr,Venikoudis:2021oee,Venikoudis:2021irr}. The theory involves a scalar field which is minimally coupled with the Ricci scalar and a function $f(\mathcal{G})$ solely depending from the Gauss-Bonnet topological invariant $\mathcal{G}$, representing higher curvature corrections and assists the accelerating expansion. For the sake of simplicity, no coupling between the scalar field and the Gauss-Bonnet topological invariant is assumed in the present article. Combining both a scalar field and a tensor Gauss-Bonnet model has the advantage that a unified description between the early and late time may be achieved. One plausible scenario could involve a dominant scalar field during the inflationary era where one for instance can implement the usual prescription of slow-roll for potential driven inflation while in the late era the extra curvature corrections, if introduced appropriately, can drive the accelerated expansion of the universe. Under this assumption, we shall examine two models of interest, the former involves a scalar field with a canonical kinetic term and scalar potential, while the latter involves a k-essence scalar field with a combination of linear and higher order kinetic term in the absence of scalar potential for simplicity. In both cases the function $f(\mathcal{G})$ has non linear terms that effectively drive the late-time. The motivation for the investigation of theories with k-essence scalar field is based on the recent striking observations from GW170817 \cite{LIGOScientific:2017ync}. The validity of these theories has been ascertained, since they can provide viable phenomenological models in primordial and late-time era \cite{Odintsov:2020qyw,Fronimos:2021czc}.  

An important feature of the present theory is that it can be compatible with latest observations and replicate the $\Lambda$CDM model. While the latter is showcased in the present article, a thorough study of tensor perturbations is a quite challenging topic and is thus not available, in principle however tensor perturbations should be quite similar to Ref. \cite{Munyeshyaka:2021vhf}. The important aspect that is worth being highlighted is that the models at hand are free of ghost instabilities given that the Ricci scalar appears in a linear form. More involved theories with terms of the form $f(R,\mathcal{G})$ are known to suffer from ghosts therefore to avoid this possibility, we shall limit our work to only $R+f(\mathcal{G})$ models and focus mainly on the late-time evolution of the universe.

Our goal of this work is to examine the phenomenology of late-time cosmological era, considering that the dark energy serves as the impact of all the involved geometric terms, in the gravitational action, something which is highlighted subsequently. Specifically, we shall examine the effect of the scalar field, which serves as low-energy quantum corrections assisted by the function $f(\mathcal{G})$. Our analysis is based on Ref. \cite{Odintsov:2021nim,Odintsov:2020vjb} where similar procedures have been performed. First of all, it is convenient to apply two variable changes in Friedmann and continuity equation of the scalar field, in order to proceed with the numerical solution of these equations. The first change involves the replacement of the cosmic time with redshift. Specifically, we introduce a differential operator which connects the cosmic time with redshift given that the Hubble's parameters is a function of redshift. As a result, the involved physical quantities are expressed in terms of redshift which yields a more manageable set of equations that need to be solved. Furthermore, we introduce an auxiliary statefinder function which is connected with the density of the dark energy to male the Friedmann equation dimensionless. By utilizing the solutions for the statefinder function and the scalar field one can extract the current numerical values for several statefinder functions in order to properly study the dynamics of the universe and make comparisons with the $\Lambda$CDM, but also trace back their evolution near the matter dominated era. In the present article, we shall focus on several statefinder parameters for the sake of consistency, namely the deceleration parameter, the jerk, the snap and parameter Om as well as the equation of state of dark energy and the density parameter of dark energy $\Omega_{DE}$. The last step in order to ascertain the validity of each model. As showcased, the models can be made to replicate the effects of $\Lambda$CDM while an infinitesimal evolution with respect to redshift is showcased. 

This paper is organized as follows: In section II  the theoretical framework of late-time Cosmology of scalar-$f(G)$ gravity is presented thoroughly as mentioned before for the first model of interest. In section III we discuss the validity of a model defining the function $f(\mathcal{G})$ and the scalar potential. In sections IV and V it is demonstrated briefly the cosmological evolution in the presence of a k-essence scalar field and $f(\mathcal{G})$ in the absence of scalar potential,
as it is usual when the theory involves higher order kinetic terms. Finally, the conclusions of our work follow at the end of the paper.

\section{late-time cosmology of scalar - $f(\mathcal{G})$ gravity}
We commence our study by properly specifying the gravitational action. The first proposed model involves 
the case of $f(\mathcal{G})$ gravity accompanied by a canonical scalar field, corresponding to the following action,
\begin{widetext}
\begin{equation}
\centering
\label{action}
S=\int{d^4x\sqrt{-g}\left(\frac{R}{2\kappa^2}-\frac{1}{2}\nabla_\mu\phi\nabla^\mu\phi-V(\phi)+\frac{f(\mathcal{G})}{2}+\mathcal{L}_{m}\right)}\, ,
\end{equation}
\end{widetext}
where R denotes the Ricci scalar, $\mathcal{L}_{m}$ represents the Lagrangian density of the perfect fluids and $f(\mathcal{G})$ is an arbitrary function for the time being which depends solely on the Gauss-Bonnet topological invariant $\mathcal{G}$. Furthermore, g is the determinant of the metric tensor $g^{\mu\nu}$, $\kappa=\frac{1}{M_P}$ is the gravitational constant while, $M_P$ denotes the reduced Planck mass. In addition, the terms $\frac{1}{2}g^{\mu\nu}\nabla_{\mu}\phi\nabla_{\nu}\phi$ and $V(\phi)$ correspond to the kinetic term of the scalar field and the scalar potential respectively. The Gauss-Bonnet topological invariant is defined as a combination of higher order curvature terms namely,
$\mathcal{G}=R_{\mu\nu\sigma\rho}R^{\mu\nu\sigma\rho}-4R_{\mu\nu}R^{\mu\nu}+R^2$ with $R_{\mu\nu\sigma\rho}$, being the Riemann curvature tensor,
while $R_{\mu\nu}$ and $R$ represents the Ricci tensor and scalar respectively. Greek indices run from $0$ to $3$ whereas Latin describes only the spatial components. As mentioned before the  $f(G)$ function can not has a linear form as $\int{d^4x\sqrt{-g}\mathcal{G}}=0$ since it is a total derivative. The case of keeping a linear term by means of non minimal coupling with the scalar field, although interesting and quite powerful for both the inflationary and the late era, it shall not be considered in the present article. Throughout our analysis, we shall assume that the geometric background corresponds to that of a flat Friedman-Robertson-Walker (FRW) metric, meaning that the line element has the following form,
\begin{equation}
\centering
\label{metric}
ds^2=-dt^2+a(t)^2\delta_{ij}dx^idx^j,\,
\end{equation}
where $a(t)$ denotes the scale factor of the Universe and the metric tensor has the form of $g_{\mu\nu}=diag(-1, a(t)^2, a(t)^2, a(t)^2)$.
 Due to the fact that the metric is assumed to be flat, the curvature terms, being the Ricci scalar and Gauss-Bonnet topological invariant, are written explicitly in terms of Hubble's parameter $H=\frac{\dot a}{a}$. Specifically, $R=6(2H^2+\dot H)$ and $\mathcal{G}=24H^2(H^2+\dot H)$ where the dot as usual implies differentiation with respect to cosmic time $t$. Furthermore, we shall assume that the aforementioned scalar field is homogeneous so as to simplify our study. This is a reasonable assumption which leads to the kinetic term being written as $X=-\frac{1}{2}\dot\phi^2$. Therefore, by implementing the variation principle in the gravitational action with respect to the metric tensor $g^{\mu\nu}$ and the scalar field $\phi$, the field equations for gravity and the continuity equation for the scalar field can be extracted. The aforementioned equations are,
\begin{equation}
\centering
\label{motion1a}
\frac{3H^2}{\kappa^2}=\frac{1}{2}\dot\phi^2+V(\phi)+\frac{\mathcal{G}f_{\mathcal{G}}-f}{2}-12\dot f_{\mathcal{G}}H^3+\rho_m\, ,
\end{equation}
\begin{equation}
\centering
\label{motion2a}
-\frac{2\dot H}{\kappa^2}=\dot \phi^2+4H^2(\ddot f_{\mathcal{G}}-H\dot f_\mathcal{G})+8\dot f_{\mathcal{G}}H\dot H+\rho_m+P_m\, ,
\end{equation}
\begin{equation}
\centering
\label{motion3a}
\ddot\phi+3H\dot\phi+\frac{dV}{d\phi}=0 ,\
\end{equation}
where $\rho_m$ and $P_m$ express the density and pressure corresponding on matter. The matter term is considered as the total contribution from non-relativistic matter namely, baryons, leptons and Cold Dark matter being the most dominant contribution while relativistic matter contains photons and neutrinos. As usual, matter can be described as a perfect fluid for any cosmological era of interest. In late-time cosmological era, the perfect fluid consists of Cold Dark Matter and radiation hence, the corresponding density is given by the following equation,
\begin{equation}
\centering
\rho_m=\rho_{nr}^{(0)}\frac{1}{\alpha^3}\left(1+\frac{\chi}{\alpha}\right),
\end{equation}
where $\chi=\frac{\rho_r^{(0)}}{\rho_{nr}^{(0)}}$. The term $\rho_{r}^{(0)}$ represents the present density of relativistic matter while, $\rho_{nr}^{(0)}$ being the current density of non relativistic matter. Such parameters shall be specified subsequently in the numerical studies. In addition, the pressure of the perfect fluid is defined as,
\begin{equation}
P=\sum_i \omega_i \rho_i,
\end{equation}
where $\omega_i$ is the equation of state parameter for both relativistic and non-relativistic matter. Here, barotropic matter is considered where $\omega_{nr}=0$ and $\omega_r=\frac{1}{3}$. The continuity equation for every component of matter has the following simple form,
\begin{equation}
\dot \rho_i+3H\rho_i(1+\omega_i)=0,
\end{equation}
where as shown, no mixing between matter components is observed. A convenient way for the investigation of the dynamics of late-time Universe is the replacement of the cosmic time with the redshift as a dynamical parameter. Redshift is defined as following,
\begin{equation}
\centering
\label{redshift}
1+z=\frac{1}{a(t)}\, ,
\end{equation}
where the current scale factor is considered to be equal to unity for simplicity. Hence, the present value of the redshift is equal to zero. According to the previous variable change, the involved time derivatives can be expressed in terms of the redshift introducing the following differential operator,
\begin{equation}
\centering
\label{diffop}
\frac{d}{dt}=-H(z)(1+z)\frac{d}{dz}\, .
\end{equation}
Based on the aforementioned assumption, the gravitational terms in the equations of motion and the time derivatives of the scalar field can be rewritten as:
\begin{equation}
\dot H=-H(1+z)H',
\end{equation}
\begin{equation}
R=12H^2-6H(1+z)H',
\end{equation}
\begin{equation}
\mathcal{G}=24H^2(H^2-H(1+z)H'),
\end{equation}
\begin{equation}
\dot{\mathcal{G}}=24(1+z)^2 H^3\left(3H'^{2}+HH''-\frac{3HH'}{1+z}\right),
\end{equation}
\begin{equation}
\label{fdot}
\dot\phi=-H(1+z)\phi',
\end{equation}
\begin{equation}
\ddot\phi=H^2(1+z)^2\phi''+H^2(1+z)\phi'+HH'(1+z)^2\phi' ,
\end{equation}
where the prime denotes the differentiation with respect to the redshift. Furthermore, for the time being, in order to investigate the dynamics of the cosmological evolution the Hubble rate and its derivatives are replaced by the dimensionless statefinder $y_H$  which is defined as,
\begin{equation}
\centering
\label{yh}
y_H(z)=\frac{\rho_{DE}}{\rho_{nr}^{(0)}}\, ,
\end{equation}
where $\rho_{DE}$ is the dark energy density which involves all the geometric terms appearing on the right hand side of Friedmann's equation hence, it is considered as,
\begin{equation}
\centering
\label{DEdensity}
\rho_{DE}=\frac{1}{2}\dot\phi^2+V(\phi)+\frac{\mathcal{G}f_{\mathcal{G}}-f}{2}-12\dot f_{\mathcal{G}}H^3\, ,
\end{equation}
while according to Raychaudhuri equation, the pressure which originates from the dark energy is:
\begin{equation}
\label{DEpressure}
P_{DE}=\dot \phi^2+4H^2(\ddot f_{\mathcal{G}}-H\dot f_\mathcal{G})+8\dot f_{\mathcal{G}}H\dot H-\rho_{DE}.
\end{equation}
This designation of dark energy density and pressure is performed in order to rewrite the Friedmann and Raychaudhuri equations in the usual form as showcased below,
\begin{equation}
\centering
\label{motion4a}
3H^2=\kappa^2\left(\rho_m+\rho_{DE}\right)\, ,
\end{equation}
\begin{equation}
\centering
\label{motion5a}
-2\dot H=\kappa^2\left(\rho_m+P_m+\rho_{DE}+P_{DE}\right).
\end{equation}
Obviously, instead of a cosmological constant now a function of redshift appears in both equations, therefore is rightfully deserves the name dark energy as it can be a formal definition of it. Another interesting conclusion which can easily be inferred from equations (\ref{motion4a}) and (\ref{motion5a}) is that dark energy must behaves as a perfect fluid since Hubble and matter components are also considered as perfect fluids. As a result the continuity equation of dark energy has the usual form,
\begin{equation}
\centering
\label{continuityDE}
\dot\rho_{DE}+3H(\rho_{DE}+P_{DE})=0,
\end{equation}
where as was the case with the matter components, no coupling between dark energy and matter is present. Note also that a constant dark energy density, meaning the cosmological constant $\Lambda$, requires an equation of state $\omega_{DE}=-1$. Therefore, one could extract similar results with the $\Lambda$CDM model by producing a dark energy equation of state which is approximately equal to $-1$ in the late era, thus requiring it to evolve infinitesimally for small redshifts, however it could change in principle drastically in previous cosmological eras of interest. In the end, only a numerical solution of the Friedmann equation can answer this but at the very least compatibility with data can be achieved. 

At this stage, let us define the statefinder function in terms of redshift as follows,
\begin{equation}
\centering
\label{yH}
y_H=\frac{H^2}{m_s^2}-\frac{\rho_{DE}}{\rho_{nr}^{(0)}}\, ,
\end{equation}
where the physical quantity $m_s^2$ is the mass scale, which is given by the following expression, $m_s^2=\frac{\kappa^2 \rho_{nr}^{(0)}}{3}=H_0^2\Omega^{(0)}_m$. According to the latest Planck data the present value of the Hubble's parameter is $H_0=67.4 \pm 0.5 \frac{km}{Mpc\times sec}$ while, the current value of the matter density parameter is $\Omega^{(0)}_{m}=0.3153$. Here, the value based on the CMB shall be used but in general one can follow the same steps using the Cepheid value and obtain different results. Based on the statefinder function, the Hubble's parameter and its derivatives can be written as,
\begin{equation}
\centering
\label{H1}
H^2=m_s^2\left(y_H+\frac{\rho_m}{\rho_{nr}^{(0)}}\right)\, ,
\end{equation}
\begin{equation}
\centering
\label{H'}
HH'=\frac{m_s^2}{2}\left(y_H'+\frac{\rho_m'}{\rho_{nr}^{(0)}}\right)\, ,
\end{equation}
\begin{equation}
\centering
\label{H''}
H'^2+HH''=\frac{m_s^2}{2}\left(y_H''+\frac{\rho_m''}{\rho_{nr}^{(0)}}\right).
\end{equation}
These terms are involved in the expressions of the Ricci and Gauss-Bonnet topological invariant equations. Although the individual expressions seem to be quite perplexed, the Friedmann equation can easily be solved numerically. In the following, we present the current matter density in terms of redshift which reads,
\begin{equation}
\centering
\label{density}
\rho=\rho_{nr}^{(0)}(1+z)^3(1+\chi(1+z)).
\end{equation}

One can ascertain whether a particular model is in agreement with the standard cosmological model $\Lambda$CDM by comparing of the numerical values of the the extra statefinder functions. These statefinder functions are, the deceleration parameter $q$, the jerk $j$, the snap $s$ and $Om$ which are given by the following expressions,
\begin{align}
\centering
\label{statefinders}
q&=-1-\frac{\dot H}{H^2}&j&=\frac{\ddot H}{H^3}-3q-2\nonumber\\
s&=\frac{j-1}{3\left(q-\frac{1}{2}\right)}&Om&=\frac{\left(\frac{H}{H_0}\right)^2-1}{(1+z)^3-1}.
\end{align}
where $H_0$ in this context should be considered as the current value of the Hubble rate as derived from the numerical analysis of a certain model and not the parameter taken from the CMB. From a more detailed perspective, the deceleration parameter and the jerk have the following forms in terms of redshift,
\begin{equation}
\centering
\label{q}
q(z)=-1+\frac{(1+z)H'}{H}\, ,
\end{equation}
\begin{equation}
\centering
\label{j}
j=(1+z)^2\left(\left(\frac{H'}{H}\right)^2+\frac{H''}{H}-\frac{2H'}{(1+z)H}\right)+1\, .
\end{equation}
For the case of constant dark energy density, the jerk is trivially equal to unity in the late era while snap should be zero by definition. Reproducing the results of the $\Lambda$CDM suggests that the jerk and the snap should evolve infinitesimally around the aforementioned values. Due to the numerical solution and the definition of the snap, a more complex form should be expected and is indeed derived as shown in the following sections. In addition, $Om$ should currently have a value close to the matter density parameter $\Omega_m$ for the sake of consistency.
Finally, the equation of state parameter $\omega_{DE}$ and the density parameter of the dark energy can be expressed in terms of the auxiliary function $y_H$. These equations read,
\begin{equation}
\omega_{DE}=-1+\frac{1+z}{3}\frac{dlny_H}{dz},
\end{equation}
\begin{equation}
\Omega_{DE}=\frac{y_H}{y_H+\rho'},
\end{equation}
where $\rho'=\frac{\rho}{\rho^{(0)}_{nr}}$. Taking into consideration the above equations, the validity of the model requires the numerical values of the aforementioned parameters for $z=0$ to be consistent with their current observational values. If this happens then, the model at hand can be in agreement with the $\Lambda$CDM. For simplicity, the area of study shall focus on te area of [-0.9,10] for redshift in order to test the model up to the matter dominated era and also make certain predictions for the evolution of the universe and see whether the description matches our current understanding.

\section{Late-time dynamics of scalar-$f(\mathcal{G})$ gravity}
In this section the late-time phenomenology of a given scalar-tensor model is investigated. The function $f(\mathcal{G})$ is defined as,
\begin{equation}
\centering
\label{f}
f(\mathcal{G})=\frac{\alpha_1}{\mathcal{G}^2}+\alpha_2\sqrt{\mathcal{G}}\, ,
\end{equation}
while the scalar potential has the following form,
\begin{equation}
\centering
\label{V}
V(\phi)=\Lambda^2(1-e^{-\sqrt{\frac{2}{3}}\frac{\phi}{M_P}})\, ,
\end{equation} 
where, $\Lambda$ denotes the cosmological constant. Parameters $\alpha_i$ are used for dimensional purposes and take the values $\alpha_1=1ev^{12}$ and $\alpha_2=1 ev^2$. Here, two separate contributions are considered, one being singular however both drive the late time. The scalar potential is used for the inflationary era. For the scalar potential, an exponential function is considered. In order to proceed with the overall phenomenology the numerical values of the free parameters must be determined. Hereafter, the cosmological constant, the Hubble's rate and the mass Planck take the following numerical values $\Lambda=11.895\cdot10^{-67}ev^2$, $H_0=1.37187 \cdot 10^{-33}ev$, $M_P=1.2\cdot 10^{28}ev$ respectively. It is worth to remind the reader that the present value of the current Hubble rate is nothing but the Hubble rate taken from CMB data and expressed in units of energy.

\begin{figure}[h!]
\centering
\label{plot2}
\includegraphics[width=17pc]{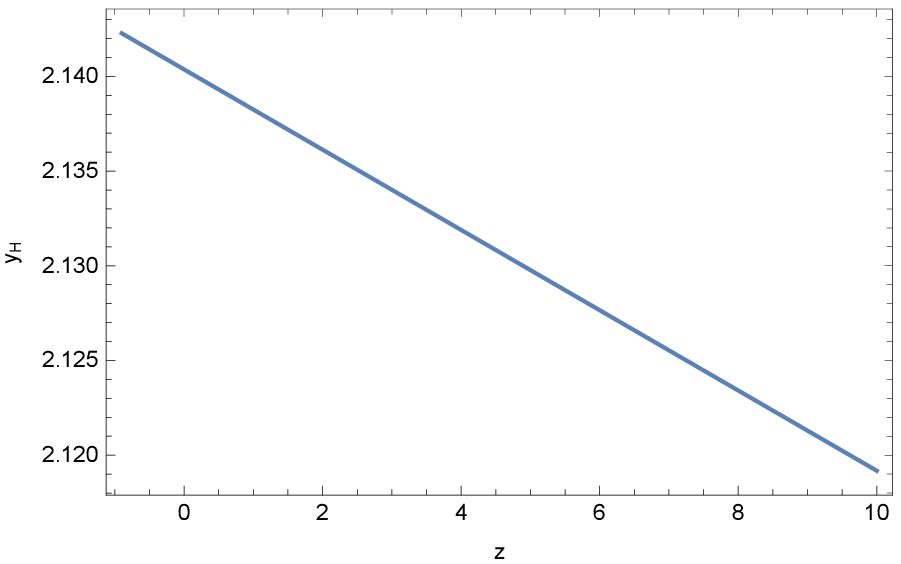}
\includegraphics[width=18pc]{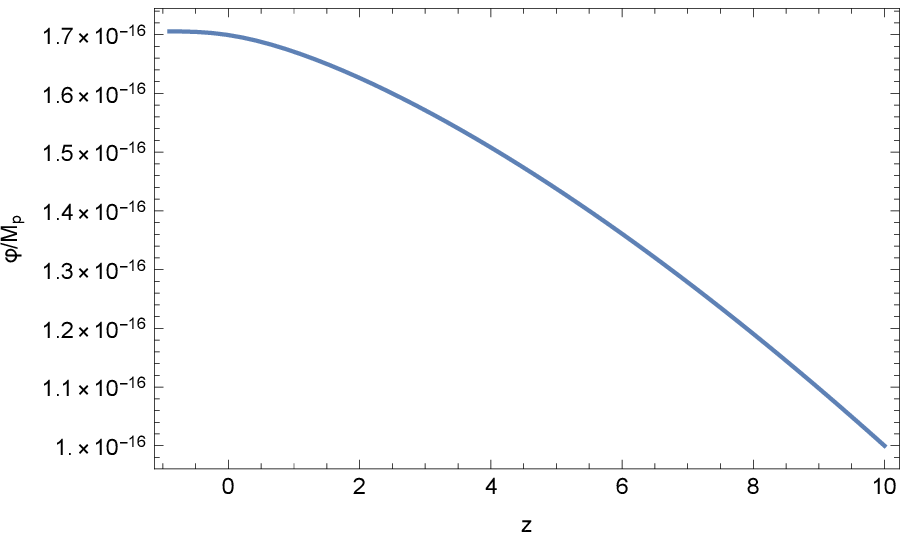}
\caption{Numerical solutions for the statefinder and the scalar field with respect to the Planck mass in the area [-0.9,10]. As shown, both are increasing with time.}
\end{figure}

\begin{figure}[h!]
\centering
\label{plot2}
\includegraphics[width=18pc]{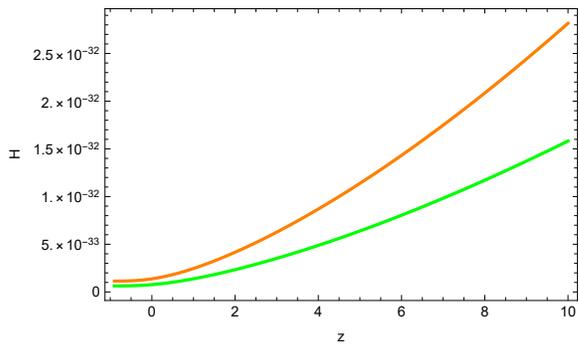}

\caption{Evolution of Hubble's parameter in the context of scalar-$f(\mathcal{G})$ gravity (orange curve) compared with the $\Lambda$CDM model (green curve) for the interval [-0.9,10]. In the aforementioned area of integration, the two functions seem to be in a relatively good agreement.}
\end{figure}

\begin{figure}[h!]
\centering
\label{plot2}
\includegraphics[width=17pc]{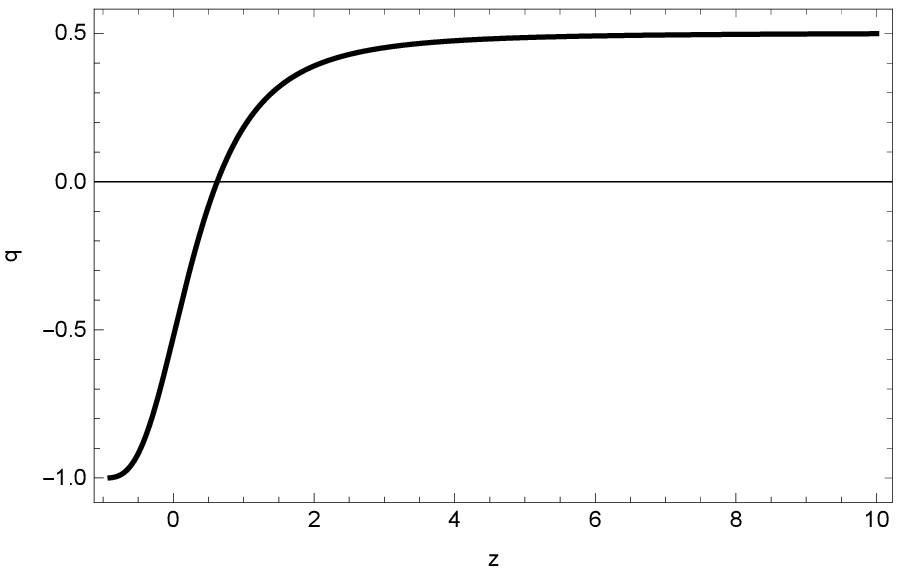}
\includegraphics[width=17.5pc]{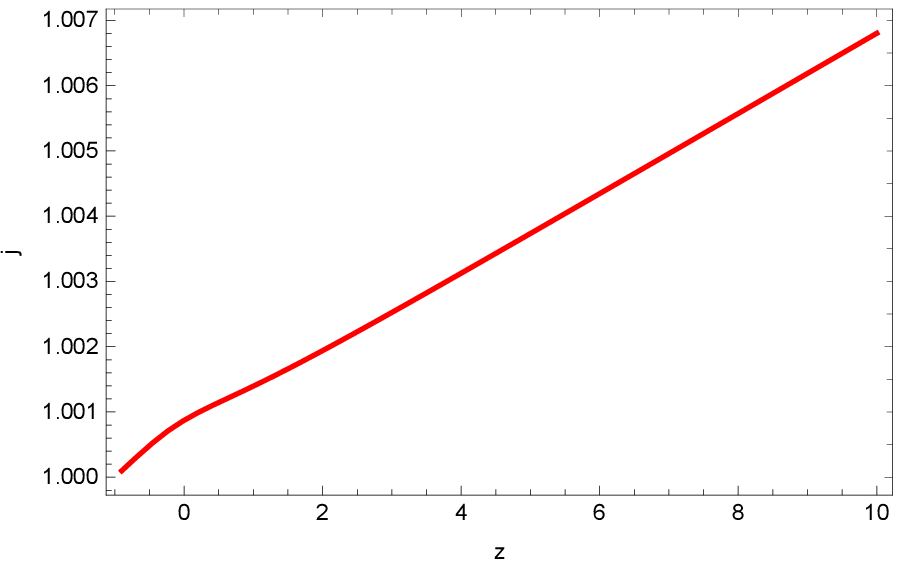}
\includegraphics[width=17pc]{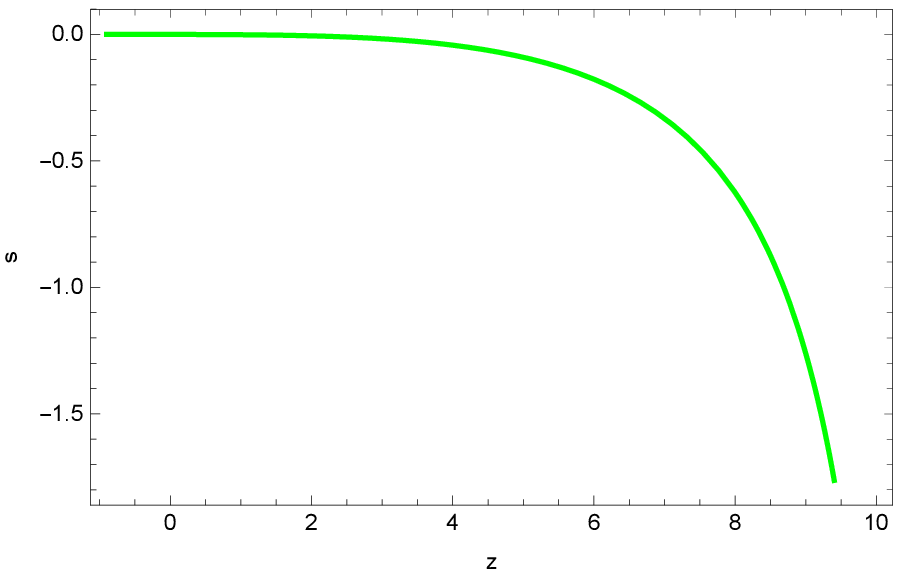}
\includegraphics[width=17.5pc]{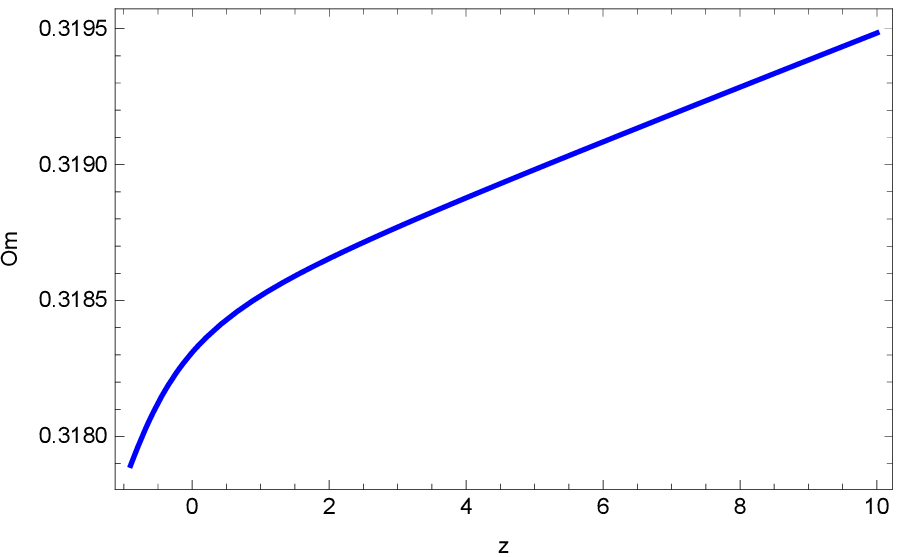}
\caption{Deceleration parameter (black), the jerk (red), the snap (green) and $Om$ (blue) as functions of redshift.}
\end{figure}
The first step of our analysis is to set the initial conditions, which are necessary in order to solve numerically the system of differential equations of motion (\ref{motion1a}-\ref{motion3a}) with respect to $y_H(z)$ and $\phi(z)$.
For this purpose, the redshift interval has been chosen as follows z=[-0.9,10] where, the initial value of redshift is considered as $z=10$. According to this assumption, the initial value of statefinder function and its first derivative with respect to redshift are given by the expressions,
\begin{equation}
\label{statefinder}
y_H(z=10)=\frac{\Lambda}{3m_s^2}\left(1+\frac{11}{1000}\right),
\end{equation}
\begin{equation}
\label{statefinderderivative}
y_H'(z=10)=\frac{\Lambda}{3000m_s^2}.
\end{equation}
Furthermore, the initial conditions for the scalar field and its first derivative are respectively, $\phi(z=10)=10^{-16}M_P$,  $\phi'(z=10)=-10^{-17}M_P$. 

At this point let us justify the dimensional analysis which is used for the involved physical quantities in cosmological evolution. According to the equation (\ref{fdot}) the first time derivative of the scalar field has dimensions [$\dot \phi ]=ev^2$, while the Hubble parameter has dimensions $[H]=ev$. This assumption implies that the initial value of the time derivative with respect to the redshift has ev units.

Given the aforementioned initial conditions, the numerical solution of the system of differential equations of motion leads to plots which are presented in Fig. 1. It can be interpreted easily that as the redshift increases, the numerical values of the statefinder $y_H$ and of the scalar field decrease. Additionally in Fig. 2 we present the evolution of the Hubble's parameter in the context of scalar-$f(\mathcal{G})$ gravity compared with the $\Lambda$ model. According to the following equation,
\begin{equation}
H_{\Lambda}(z)=H_0\sqrt{\Omega_{\Lambda}+\Omega_{nr}(1+z)^3+\Omega_{r}(1+z)^4},
\end{equation}
the Hubble's parameter depends on the value of redshift and on the density parameters $\Omega_{\Lambda}$, $\Omega_{nr}$ and $\Omega_{r}$ for any cosmological era of interest. In late-time era the numerical values of the aforementioned parameters are $\Omega_{\Lambda}\simeq0.681369$ and $\Omega_{nr}\simeq 0.3153$ based on the latest Planck data \cite{Planck:2018vyg} while $\Omega_r$ is irrelevant, as suggested by $\chi$ which is of order $\mathcal{O}(10^{-4})$.
\begin{table}
\centering
\begin{tabular}{|c|c|c|}\hline
Statefinders&Numerical Results& $\Lambda$CDM Value\\ \hline
q(0)&-0.522536&-0.535\\ \hline
j(0)&1.00087&1\\ \hline
s(0)&-0.000284313&0\\ \hline
Om(0)&0.318309&0.3153$\pm$0.007\\ \hline
$\omega_{DE}(0)$&-1.00033&-1.018$\pm$0.31\\ \hline
$\Omega_{DE}(0)$&0.681499&0.6847$\pm0.0073$\\ \hline
\end{tabular}
\caption{Numerical comparison of the current values of the extra statefinder functions. The current values suggest that the model is indeed in agreement with the $\Lambda$CDM.}
\end{table}
\begin{figure}
\centering
\label{plot4}
\includegraphics[width=17pc]{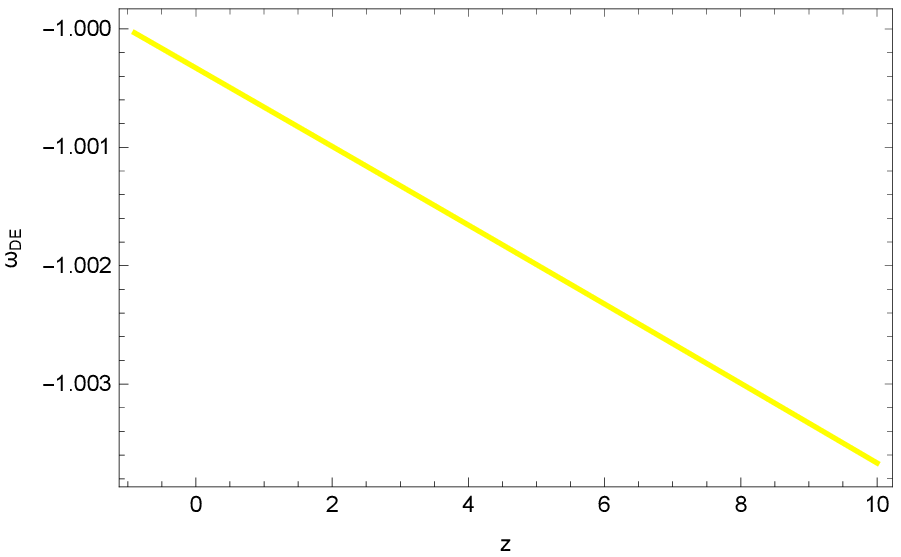}
\includegraphics[width=16.6pc]{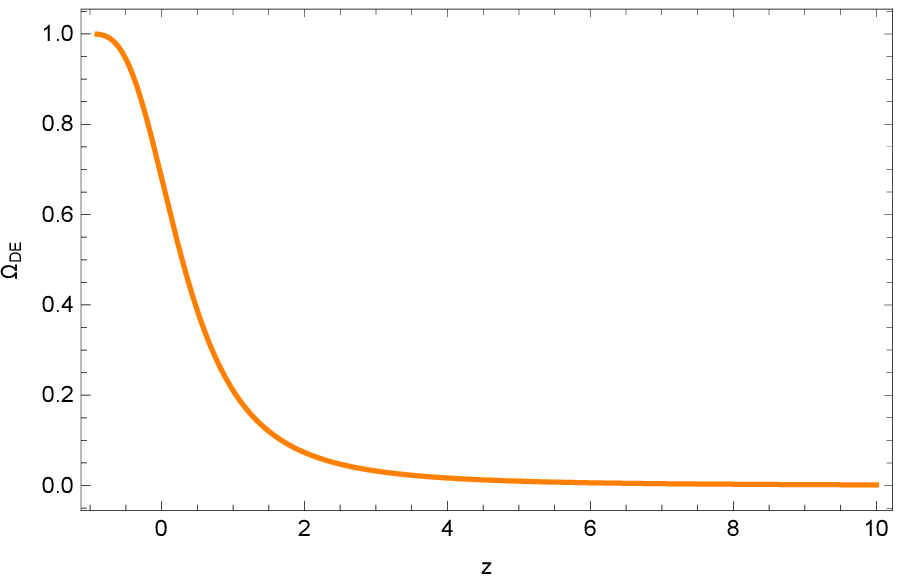}
\caption{Dark energy EoS (yellow) and density parameter (orange) as functions of redshift. While the dark energy density is smoothly increasing and reaches unity in the future, the dark energy EoS as expected is infinitesimally evolving thus rendering the model compatible with $\Lambda$CDM.}
\end{figure}
One can ascertain if a particular model of interest can be considered as viable, comparing the numerical statefinder parameters and the dark energy parameters with the $\Lambda$CDM model. In the context of scalar-$f(\mathcal{G})$ gravity the current value of the deceleration parameter is $q(0)\simeq-0.52$ which indicates good agreement with the $\Lambda$CDM as presented in Table I. In addition, the jerk and the snap parameters take also numerical values which are
in consistency with the $\Lambda$CDM as the former is almost equal to unity, while the latter can be considered approximately as zero. The Om statefinder parameter which indicates the present value of matter density is almost equal to $\Omega_m$ as expected. The dark energy density today is predicted $\Omega_{DE}(0)\simeq 0.68$ which means that the condition $\Omega_{DE}(0)+\Omega m(0)\simeq1$ is satisfied given that the radiation today can be considered that has negligible contribution in late-time cosmological era. In Figures 3 and 4 it is demonstrated the evolution of the aforementioned statefinder parameters with respect to redshift for the interval [-0.9,10]. Here, it is showcased that the deceleration parameter reaches the value of $q=0.5$ for large redshifts and in the future seems to reach the value of $q=-1$. This implies that the model smoothly unifies the late-time era with the matter dominated era while in the future a de Sitter expansion is achieved. Similarly, jerk is infinitesimally increasing however it resides near the value of $j=1$ as expected. Now, the snap, as mentioned before, seems to be currently near the expected value of $s=0$ but decreases with redshift. This decrease is a direct consequence of the infinitesimal increase of the jerk. It is expected that the snap will diverge at some value of redshift where the numerical value of the deceleration parameter shall be exactly $q=0.5$. Contrary to the snap, $Om$ seems to decrease with time while the dark energy density parameter $\Omega_{DE}$ increases. As a result, the Friedmann constraint $\Omega_{nr}+\Omega_r+\Omega_{DE}=1$ seems to be satisfied approximately by replacing the non relativistic density parameter with $Om$. Finally, the dark energy EoS does not currently coincide with the $\Lambda$CDM but in the future where dark energy will be the only dominant contribution to the Friedmann equation, the EoS shall obtain the value $\omega_{DE}=-1$ during the de Sitter expansion. Once again, the infinitesimal evolution is indicative of the compatibility of the current model with the $\Lambda$CDM.
\section{Theoretical Framework of $f(\mathcal{G})$ gravity accompanied by a k-essence scalar field} 
In the following section it is demonstrated briefly the theoretical framework of late-time cosmological evolution in the presence of a k-essence scalar field assisted by $f(G)$ gravity in the absence of scalar potential. K-essence scalar field involves a nonstandard kinetic term which consists of the combination of the usual part $X=\frac{1}{2}g^{\mu\nu}\partial_\mu\phi\partial_\nu\phi$ and a higher order kinetic term, in this case a squared contribution $X^2$ shall be assumed.
The gravitational action for such a model is written as,
\begin{widetext}
\begin{equation}
\centering
\label{action2}
S=\int{d^4x\sqrt{-g}\left(\frac{R}{2\kappa^2}-c_1 X-c_2 X^2+\frac{f(\mathcal{G})}{2}+\mathcal{L}_{matter}\right)}\, ,
\end{equation}
\end{widetext}
where the auxiliary parameters $c_1$ and $c_2$ are also arbitrary constants with the first being dimensionless while the latter has mass dimensions $[m]^{-4}$ for the sake of consistency. Here, $c_1$ is used in order to have the phantom case corresponding to $c_1=-1$ available for the reader however in the following numerical analysis only a canonical contribution shall be assumed.
By implementing the variation principle with respect to the metric tensor and the scalar field respectively, the equations of motion are derived which in this case read,
\begin{widetext}
\begin{equation}
\centering
\label{motion1b}
\frac{3H^2}{\kappa^2}=\rho_{m}+\frac{c_1}{2}\dot\phi^2-\frac{3c_2}{4}\dot\phi^4+\frac{\mathcal{G}f_{\mathcal{G}}-f}{2}-12\dot f_{\mathcal{G}}H^3\, ,
\end{equation}
\begin{equation}
\centering
\label{motion2b}
-\frac{2\dot H}{\kappa^2}=\rho_{m}+P_{m}+2c_1\dot\phi^2-4c_2\kappa^4\dot\phi^4+4H^2(\ddot f_{\mathcal{G}}-H\dot f_\mathcal{G})+8\dot f_{\mathcal{G}}H\dot H\, ,
\end{equation}
\begin{equation}
\centering
\label{motion3b}
c_1(\ddot\phi+3H\dot\phi)-3c_2(\ddot\phi+H\dot\phi)\kappa^4\dot\phi^2=0\, ,
\end{equation}
\end{widetext}
Our methodology for the investigation of the late-time era is based on the analysis which is performed in Section II. However, as the system of equations of motion has been altered due to the inclusion of the quadratic kinetic term $X^2$, the density and the pressure of the dark energy are given now by the following relations respectively,
\begin{widetext}
\begin{equation}
\centering
\label{DEdensity2}
\rho_{DE}=\frac{c_1}{2}\dot\phi^2-\frac{3c_2}{4}\dot\phi^4+\frac{\mathcal{G}f_{\mathcal{G}}-f}{2}-12\dot f_{\mathcal{G}}H^3,
\end{equation}
\begin{equation}
\centering
\label{DEpressure2}
P_{DE}=2c_1\dot\phi^2-4c_2\kappa^4\dot\phi^4+4H^2(\ddot f_{\mathcal{G}}-H\dot f_\mathcal{G})+8\dot f_{\mathcal{G}}H\dot H-\rho_{DE}.
\end{equation}
\end{widetext}
The rest expressions remain unaffected by the new inclusion in the gravitational action. This is really convenient as one can use exactly the same equations as before for a variety of gravitational actions and just specify properly what is dark matter density and pressure in this context. Compatibility with observations, if it can be achieved, is studied by solving the equations of motion numerically no matter the inclusion of additional terms in the action.

\section{Numerical Analysis and Validity of the model}
We present the late-time phenomenology in the presence of a k-essence scalar field assisted by $f(\mathcal{G})$ gravity. For the shake of simplicity the case of absent scalar potential is considered, which is a usual assumption for noncanonical scalar fields with higher order kinetic terms. First of all, the arbitrary function $f(\mathcal{G})$ which appears in the gravitational action (\ref{action2}), is now specified as follows,
\begin{equation}
f(\mathcal{G})=\frac{\alpha_1}{\mathcal{G}}+\alpha_2\mathcal{G}^{1/4}
\end{equation}
where $\alpha_1$ and $\alpha_2$ are auxiliary parameters introduced for dimensional purposes but are both replaced with the value $\alpha_i=1$ for simplicity. Once again, our analysis begins from the imposed initial conditions. For this reason it is assumed that the statefinder function as well as its first derivative with respect to redshift are given by the following expressions $y_H(z=10)=\frac{1.05\Lambda}{3m_s^2}$ , $y_H'(z=10)=\frac{\Lambda}{3000m_s^2}$ respectively. Furthermore, the initial value of the scalar field and its first derivative is, $\phi(z=10)=10^{-16}M_P$ and $\phi'(z=10)=-10^{-17}M_P$. In addition, the cosmological constant and the Planck mass remain exactly the same as in the previous model.
The numerical solution of the system of the differential equations (\ref{motion1b}-\ref{motion3b}) with respect to the statefinder function is presented in Figure 5. The current value of the statefinder function is $y_H(0)=2.20394$. According to the plot it is abundantly clear the absence of dark energy oscillations during the cosmological evolution.
\begin{figure}
\centering
\label{plot5}
\includegraphics[width=17pc]{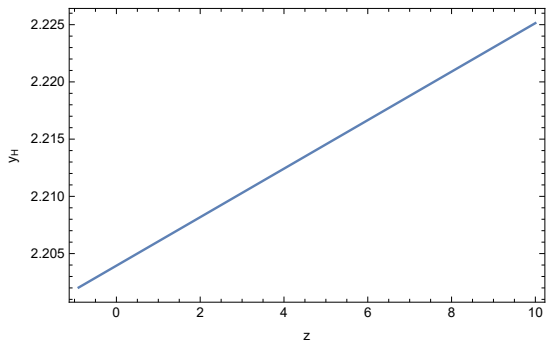}
\caption{Numerical solution of the statefinder function for the interval [-0.9,10]. In contrast to the previous model, the statefinder function decreases with time.}
\end{figure}
In this case, the solutions of the equations of motion seem to decrease with time compared to the previous model. The difference lies with the initial condition for statefinder $y_H$ for the current model which as shown manages to alter the properties of dark energy. In Table II, the present numerical values of the statefinder functions of the given model are demonstrated and as shown are in consistency with the $\Lambda$CDM standard cosmological model.
\begin{figure}[h!]
\centering
\label{plot2}
\includegraphics[width=17pc]{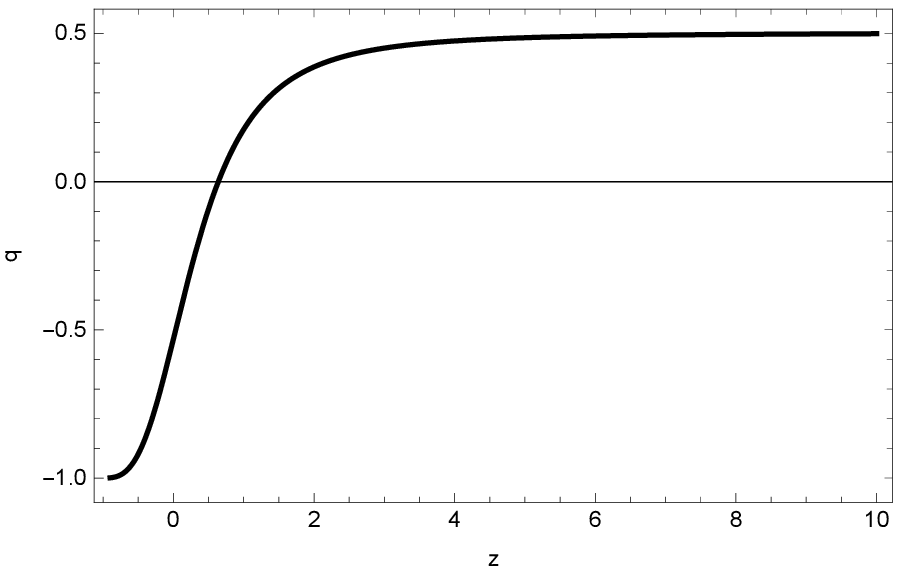}
\includegraphics[width=17.5pc]{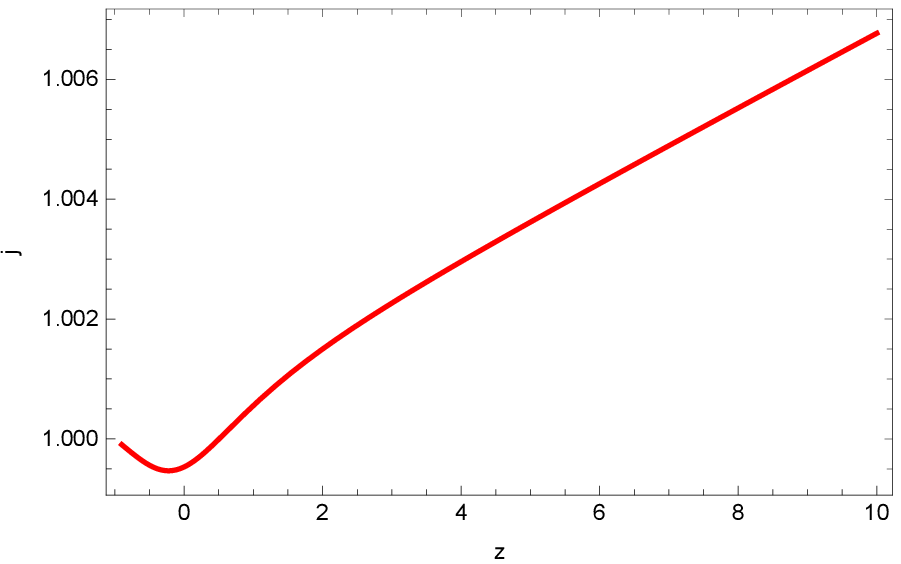}
\includegraphics[width=17pc]{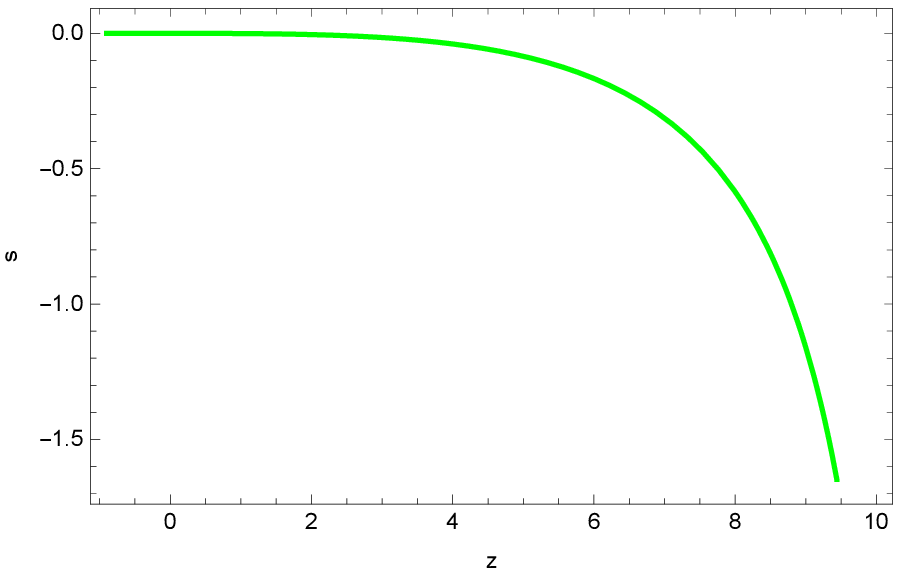}
\includegraphics[width=17.5pc]{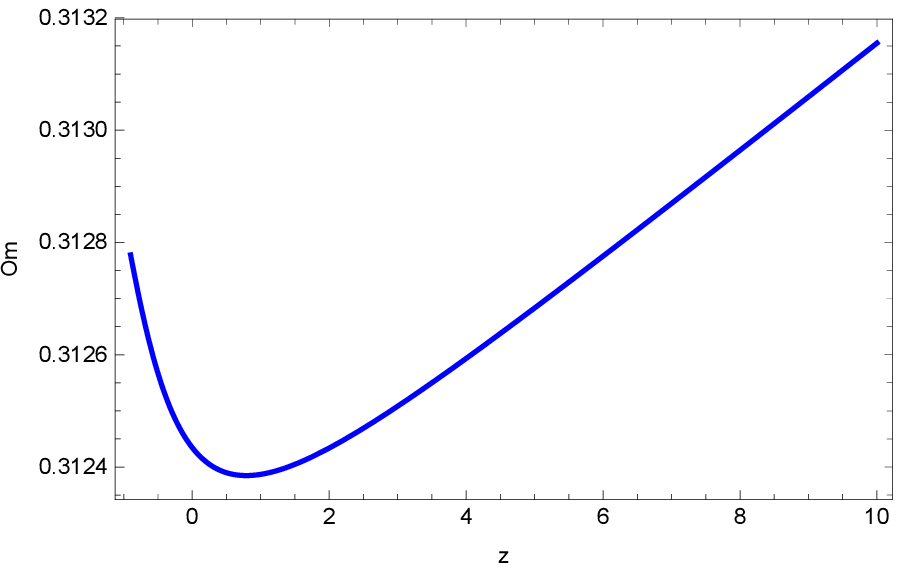}
\caption{Deceleration parameter (black), the jerk (red), the snap (green) and $Om$ (blue) as functions of redshift in the late-time era. The overall behaviour seems quite similar between the models however differences in parameters containing higher derivatives of Hubble are spotted for small values of redshift.}
\end{figure}
\begin{figure}
\centering
\label{plot6}
\includegraphics[width=17pc]{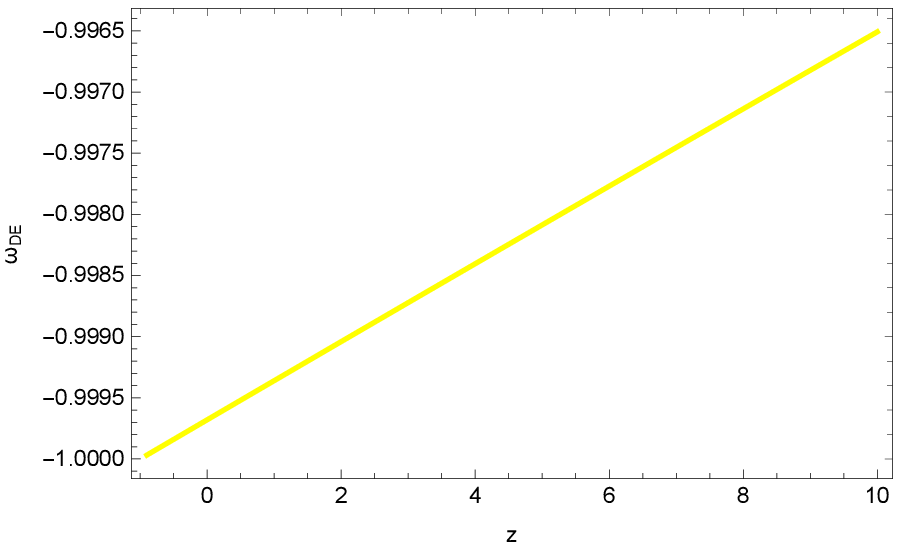}
\includegraphics[width=16.6pc]{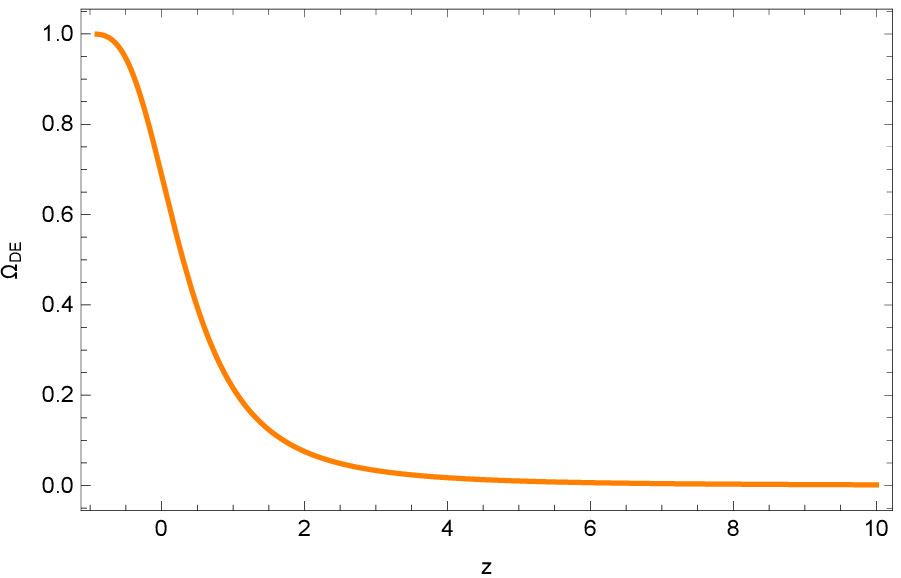}
\caption{Dark energy EoS (yellow) and density parameter (orange) as functions of redshift. In the current model, the dark energy EoS is decreasing for large values of redshift.}
\end{figure}
\begin{table}
\centering
\begin{tabular}{|c|c|c|}\hline
Statefinders&Numerical Results& $\Lambda$CDM Value\\ \hline
q(0)&-0.531348&-0.535\\ \hline
j(0)&0.999532&1\\ \hline
s(0)&0.000151217&0\\ \hline
Om(0)&0.312435&0.3153$\pm$0.007\\ \hline
$\omega_{DE}(0)$&-0.999679&-1.018$\pm$0.31\\ \hline
$\Omega_{DE}(0)$&0.687818&0.6847$\pm0.0073$\\ \hline
\end{tabular}
\caption{Numerical values of the statefinder parameters in k-essence-$f(\mathcal{G})$ gravity compared with the $\Lambda$CDM mdoel.}
\end{table}
Here, the deceleration parameter seems to behave exactly as the previous model. This would be expected since the form of $f(\mathcal{G})$ in the current model shares similar characteristics as the previous, meaning that both have a singular term given by an inverse power-law and a power-law part with exponent ranging in the area $0<n<1$. Due to the fact that the functions are inherently different, higher derivatives of Hubble such as the jerk are affected. In Fig.6 the jerk has a different behavior for small redshifts. The same applies to $Om$ which as shown starts increasing again in the future. The initial conditions used in both cases are quite similar therefore the different choices in models, both the $f(\mathcal{G})$ and the k-essence part over the scalar potential affect this result. The major impact seems to lie with the EoS of dark energy in Fig.7 where it behaves opposite to the previous model. Now the eoS increases with the increase of redshift with the change once again being spotted in the third decimal. Infinitesimal evolution is in an essence what manages to replicate the $\Lambda$CDM results to a good extend. The behavior of the dark energy density parameter as expected is similar to the previous case where it increases with time. Overall, while similar models are studied, different properties for dark energy may be obtained if one studies scalar field assisted $f(\mathcal{G})$ gravity models.

\section{Conclusions}
In this paper we investigated the late-time cosmological era in the context of a given scalar-tensor gravitational theory. Specifically the phenomenological implications of the inclusion of a scalar field and a function $f(\mathcal{G})$ in the cosmological evolution are examined. Two models of interest were presented, the first involved the case of a canonical scalar field, while the second a scalar field with a standard and a higher order kinetic term without a scalar potential. In order to simplify our analysis we performed two variable changes. Firstly, we replaced the cosmic time with redshift and secondly we introduced the statefinder function in terms of dark energy instead of Hubble's parameter. Based on the aforementioned transformations the system of equations of motions was extracted and it was solved numerically with to the statefinder function and of the scalar field. For both models which are investigated, it is proved that compatibility with the latest Planck data and agreement with the $\Lambda$CDM can be achieved. Comparing with cosmological models, which arise in the context of $f(R)$ gravitational theories the main benefits of the theory worth to be mentioned. In this context, dark energy oscillations during the early stages of matter dominated era are absent even though power-law models with exponents $0<n<1$ are considered. The choice of curvature invariant seems to matter. The scalar field present is mainly used in order to unify inflation with the late-time. The interesting part lies with the choice of a linear Ricci scalar, which was mainly used in order to obtain simler equations of motion and also avoid the appearance of ghost instabilities which are are typically present in $f(R,\mathcal{G})$ models. The idea of such models however which are assisted by scalar fields is quite interesting and could result in interesting characteristics for dark energy. We hope to address this interesting and rather challenging topic in a future work.

\end{document}